\documentclass[lettersize,journal]{IEEEtran}
\usepackage[colorlinks,urlcolor=blue,linkcolor=blue,citecolor=blue]{hyperref}
\usepackage{color,array}
\usepackage{listings}
\usepackage[linesnumbered,ruled,vlined]{algorithm2e}
\usepackage{algorithmic}
\usepackage{listings}
\usepackage{amsmath}
\usepackage{xspace}
\usepackage{tikz}

\usepackage{pifont}
\usepackage{url}
\usepackage{multirow}
\usepackage{caption}
\usepackage{subcaption}
\usepackage{amssymb}
\DeclareUnicodeCharacter{2061}{}
\usepackage{epstopdf}   
\usepackage{epsfig}
\usepackage{graphicx}
\usepackage{comment}
\usepackage{amsthm}

\usepackage{enumerate}
\usepackage[utf8]{inputenc}
\usepackage{lettrine}
\usepackage{hyperref}
\usepackage{array}
\usepackage{stfloats}
\usepackage{blindtext}
\usepackage{fancyhdr}
\usepackage{graphicx}
\usepackage{verbatim}
\usepackage{xcolor}
\usepackage{xurl}
\usepackage{comment}
\usepackage{lettrine}
\usepackage{cite}
\usepackage[normalem]{ulem}
\usepackage{pifont}
\begin{document}
\title{EnFed: An Energy-aware Federated Learning in Resource Constrained Environments for Human Activity Recognition}
\author{
Anwesha Mukherjee and Rajkumar Buyya, \IEEEmembership{Fellow, IEEE}

\thanks{A. Mukherjee and R. Buyya are with the Quantum Cloud Computing and
Distributed Systems (qCLOUDS) Laboratory, School of Computing and
Information Systems, University of Melbourne, Australia. (e-mail:
rbuyya@unimelb.edu.au).
A. Mukherjee is also with the Department of Computer Science, Mahishadal
Raj College, Mahishadal, West Bengal, India. (e-mail: anweshamukherjee2011@gmail.com)}
}

\maketitle

\begin{abstract}
\textcolor{black}{The human activity recognition (HAR) and recommendation applications for mobile users require a privacy-aware and accurate data analysis model with lower time and lower energy consumption. The use of federated learning (FL) to develop a privacy-aware HAR model is an emerging research topic. However, the participating mobile devices in the FL process may slow down due to their limited computational resources, connectivity interruption, and limited battery life. To address these challenges, this paper proposes an energy-efficient FL method referred to as EnFed, with a case study on HAR. In EnFed, a mobile device that needs a model for an application requests its nearby devices with respect to an incentive. The nearby devices, which agree to the offered incentive, send their local model updates, i.e., updated local model parameters for that application, to the requesting device. The device, after receiving local model updates from the contributors, aggregates them to build a global model and then fits the model to its own dataset to build its own local model. The results show that using EnFed a resource-limited device obtains an accurate prediction model at lower time and lower energy consumption. The experimental results show that EnFed achieves above 95\% prediction accuracy and outperforms the baselines. EnFed also reduces the response time above 90\% than the cloud-only framework. The comparative study shows that EnFed outperforms the existing HAR approaches.}
\end{abstract}


\section{Introduction}
Federated learning (FL) has become an emerging topic in data analytics with a focus on privacy preservation, distributed learning, and collaborative training \cite{nguyen2021federated, pavlidis2024federated}. With the huge increase in smartphone usage and the growing popularity of various activity monitoring and recognition applications, several challenges are introduced, such as a huge overhead on the cloud, the requirement of high network bandwidth, and the limited battery life of mobile devices. There are various applications which collect user data, perform analysis, and give output based on the predicted results. Human activity recognition (HAR) and recommendation applications, such as healthcare applications, activity monitoring applications, and dietary recommendation apps, collect users' personal data related to health, food habits, and activities, which are confidential. In such a case, data storage and analysis on the server may compromise the user's privacy. Furthermore, data transmission to the cloud increases network traffic and cloud overhead, and requires seamless network connectivity that may not always be available to mobile users. Furthermore, a global model may not always provide good results for personalized applications. FL can deal with these issues by providing the facility of local data analysis along with collaborative training so that finally each device can have a personalized local model along with a global model \cite{nguyen2021federated, bera2024flag}. However, centralized federated learning (CFL) also needs good network connectivity for exchanging model updates, i.e., updated model parameters with the server that works as the aggregator. In decentralized federated learning (DFL) \cite{bera2024fedchain, hashemi2021benefits}, participating devices form a network among themselves and perform a collaborative learning process to build a generalized model. 
\par
In our paper, an energy-efficient FL approach is proposed for resource-limited mobile devices with limited battery life to obtain a model for an application.
\subsection{Motivation and Contributions}
\textcolor{black}{Mobile users access various types of activity recognition and recommendation applications. The use of centralized cloud-based data analysis and storage compromises data privacy and increases cloud overhead and response time. As users' data are highly sensitive and confidential, privacy-aware data analysis is vital in HAR. Furthermore, each device requires a personalized local model and a global model. FL has addressed these issues by allowing the devices to analyze the data locally and build a global model through collaborative training without sharing raw data. However, the mobile devices have limited computational resources, limited battery life, and may not have seamless network connectivity. Hence, they may not be able to deliver the model updates on time, which affects the entire FL process. Asynchronous FL \cite{xu2023asynchronous} has allowed the aggregation after receiving model updates from a participating device without waiting for all participating devices to finish training. However, the devices may not always have sufficient network connectivity to exchange model updates with the server until a final global model is created. Though DFL \cite{mukherjee2024federated} has allowed the devices to collaborate among themselves either through a mesh topology or a ring topology (peer-to-peer network) without the coordination of a central server, all participating devices may not have sufficient battery levels to finally complete the process. Furthermore, the nearby devices may have updated models with them. Hence, an energy-efficient FL approach is required for the resource-constrained mobile devices with limited battery life and limited connectivity. \textit{The objective of this work is to address these challenges and propose an energy-efficient FL approach.} To achieve this, we make the following key contributions:}
\begin{itemize}
    \item \textcolor{black}{An \textit{En}ergy-aware \textit{Fed}erated Learning (\textit{EnFed}) method is proposed for developing a personalized local model for an application meeting a desired accuracy level, at minimal training time and minimal energy consumption of the mobile device.} 
    \item \textcolor{black}{EnFed allows individual model training with the support from nearby devices with updated models for the same application.} 
    \item \textcolor{black}{In EnFed, a resource-constrained mobile device with limited battery life builds a personalized local model for an application based on the cooperation of nearby devices with respect to an incentive.}
    \item \textcolor{black}{EnFed maintains a trade-off between the accuracy level and the training time along with energy consumption to overcome the issue of limited battery life of mobile devices without compromising the desired accuracy level. As the underlying data analysis model, we use Long Short-Term Memory (LSTM) networks and Multi-Layer Perceptrons (MLP) in EnFed because LSTM is suitable for time series data and sequential data, and MLPs can learn complex data patterns.}
    \item \textcolor{black}{The performance of \textit{EnFed} is compared with the CFL and conventional DFL-based systems, and we observe that a good prediction accuracy is achieved by \textit{EnFed} using LSTM and MLP at a lower training time and lower energy consumption of the device.}
\end{itemize}
The rest of the paper is organized as follows. Section \ref{rel} presents the existing works on FL. Section \ref{pro} demonstrates the proposed FL approach with a detailed discussion on time consumption and energy consumption. Section \ref{perf} presents the performance of EnFed. Finally, Section \ref{con} concludes the paper with future research directions.
\section{Related Work}
\label{rel}
This section discusses the existing works on FL and HAR, along with their limitations and how EnFed addresses them.
\subsection{Existing works on FL}
FL has gained its popularity for privacy-aware data analysis in various domains including Internet of Things (IoT), mobile network, etc. Integrating the advantages of CFL and DFL architectures, a hybrid FL model was proposed in \cite{liu2024fedcd}. The model was distributed based on the consensus distances and layer sizes to enhance the training speed, reduce network bandwidth pressure, and improve prediction accuracy \cite{liu2024fedcd}. In \cite{lee2021opportunistic}, an opportunistic FL approach was proposed for pervasive computing applications. 
In \cite{imteaj2021survey}, the authors demonstrated the use of FL for resource-limited IoT devices. The use of FL with mobile edge computing (MEC) was highlighted in \cite{lim2020federated}. The authors did a thorough literature survey on the existing FL solutions for MEC. In \cite{ye2020edgefed}, the authors proposed an FL-based approach for updating local models on mobile devices with the coordination of the edge server. However, coordination with the server incurs some communication overhead. For FL, not only computational latency but also communication latency plays a vital role, which was highlighted in \cite{chen2023service}. With an objective of service delay minimization, an FL approach was proposed for mobile devices \cite{chen2023service}. The authors discussed gradient quantization and weight quantization to generate a delay-aware model. Along with latency, energy is also an important parameter that needs to be considered while developing an FL-based model for mobile devices. In \cite{shi2022toward}, the authors proposed an approach to maintain a trade-off between the energy consumption for computation and communication during FL. The authors discussed gradient sparsification and quantization to reduce communication energy, and weight quantization and model pruning to reduce computing energy. Although gradient sparsification and quantization, weight quantization, and model pruning were adopted by some existing works, there is a probability of reducing the overall accuracy of the model. Based on crowdsourcing and game theory, an approach was proposed for on-device FL in \cite{pandey2020crowdsourcing}. \textcolor{black}{We observe that most of the existing literature used CFL in their approaches, where a central coordinator performs aggregation and develops the global model. However, there is a probability that the connectivity with the central server may not always be available to the mobile device. Furthermore, the devices may slow down during the FL process. Though asynchronous FL \cite{xu2023asynchronous} allows the aggregation after receiving model updates from a participating client without waiting for all clients to finish training, the devices may not always have sufficient network connectivity to exchange model updates with the server until a final global model is developed. Though DFL \cite{mukherjee2024federated} has been adopted to allow the devices to collaborate among themselves without the coordination of a central server, the mobile devices with poor battery levels may fail to complete the DFL process. \textit{EnFed has addressed these issues by allowing a mobile device with limited battery life to build a model for an application by taking cooperation from nearby devices with updated models with respect to an incentive.}}
\subsection{Existing works on HAR}
The use of machine learning (ML) and deep learning (DL) on activity recognition was explored in many research works. The use of DL for activity recognition was discussed in \cite{tang2022multiscale}. The use of personalized DL for HAR was studied in \cite{ferrari2023deep}. In \cite{priyadarshini2023human}, various types of ML and DL algorithms were used for activity recognition in cyber-physical systems. The authors explored the use of deep convolutional neural network (CNN) in \cite{kaya2024human} for activity recognition. \textcolor{black}{However, privacy protection is a vital issue in HAR especially when cloud-based data analysis takes place using ML or DL. Furthermore, response time and energy efficiency are also significant along with prediction accuracy.} In \cite{jha2022hybrid}, an accuracy and energy-aware activity recognition model was proposed. In \cite{zou2024gt}, a generic graph-based temporal framework was discussed for HAR. To build an accurate HAR model, the dataset needs sufficient number of samples for each class. A dataset with class imbalance may not generate an accurate model. In \cite{guo2021evolutionary}, a dual-ensemble class imbalance learning strategy was proposed for activity recognition. To address data privacy concerns and build an accurate model through collaborative training, FL came. In \cite{cheng2023protohar}, the authors explored the use of FL for activity recognition. FL-based health recommendation was explored in \cite{ghosh2023feel}. For location-based services, a geographical point-of-interest (POI) recommendation strategy was proposed in \cite{huang2022geographical} based on FL. In \cite{varlamis2023using}, the authors proposed an FL-based approach for recommendation generations using LSTM. For context-aware recommendations with privacy protection, FL was used in \cite{ali2021federated}. In \cite{shaik2022fedstack}, an FL-based approach was used for activity monitoring. The authors proposed FedStack which performed stack-based model predictions using FL \cite{shaik2022fedstack}. In \cite{albogamy2025federated}, FL with LSTM and gated recurrent unit (GRU) was used for activity recognition. 

\textcolor{black}{However, most of existing systems used CFL in their frameworks, which need seamless network connectivity for exchanging updates to finally build an optimized global model. Furthermore, the participating devices may slow down during the FL process due to resource limitation, which is known as the straggler effect \cite{mukherjee2025joint}. The computational resource limitation, limited battery life, and connectivity interruption create obstacles during the generation of a global model through CFL or DFL when the participating nodes are mobile devices. To provide a solution, this paper proposes an FL model where a resource-constrained device can obtain a model for an application based on the cooperation of nearby devices. A comparison between the proposed and existing FL-based recognition and recommendation systems is presented in Table \ref{tab:1}. Most of the existing approaches aimed to generate an accurate model but did not consider the energy consumption, which is a critical issue for mobile devices. Furthermore, only a few have measured the training time, which is also an important aspect when developing a model. In our work, a trade-off is maintained between the prediction accuracy and training time along with energy consumption so that a model with a desired accuracy level can be obtained at minimal time and minimal energy consumption of the device.} 

\begin{table}
    \centering
    \caption{\centering{Comparison of EnFed with existing FL-based recognition and recommendation systems}}
    \begin{tabular}{c c c c c}
        \hline
         Work & Application & Used& Measured & Measured\\
         & &FL & training time & energy\\
        \hline
        Cheng & Activity & Yes & No & No\\
      et al.  \cite{cheng2023protohar}& recognition & &  & \\
        Ghosh &  Healthcare & Yes & No & No\\
        et al. \cite{ghosh2023feel} & & & &\\
        Huang &  Geographical POI & Yes & Yes & No\\
         et al. \cite{huang2022geographical} & recommendation & & &\\
        Varlamis & Energy-efficient & Yes & Yes & No\\
         et al. \cite{varlamis2023using} &smart home & & &\\
        Ali & Context-aware & Yes & No & No\\
         et al. \cite{ali2021federated} &recommendation& & &\\
        Shaik & Activity & Yes & No & No\\
         et al. \cite{shaik2022fedstack} &recognition& & &\\
        Albogamy & Activity & Yes & No & No\\
    \cite{albogamy2025federated} &recognition& & &\\
        EnFed & Activity & Yes & Yes & Yes\\
       (proposed) & recognition &  & & \\
       \hline
    \end{tabular}
    \label{tab:1}
\end{table}

\section{EnFed:Proposed FL approach}
\label{pro}
Mobile devices such as smartphones, laptops, tablets, have limited battery life, and thus, energy consumption is a vital issue. However, users prefer to access various types of activity monitoring and recognition applications using their mobile devices. In addition, several applications require connectivity to the server to update the respective models. However, seamless network connectivity is not always available, and the device may not always have sufficient battery level to receive model updates from the server to upgrade its own model. The device also needs to install and update many applications. A situation may arise when the current battery level of a mobile device is below a minimum threshold value or connectivity to the server is not available, but an updated application model is required at that time. In the proposed FL approach, the device requests cooperation from nearby devices to initialize or update the application model, with respect to an incentive. Due to the limited battery life of the mobile device, we have considered both the accuracy level and the current battery level to decide the number of rounds to update the model. The battery discharge rate can be non-linear and depends on various factors. By the minimum battery life we refer to the minimum charge required to execute the respective application by the device. The battery threshold will vary according to the type of application, device configuration, bandwidth, etc. The mathematical notations used in our approach are summarized in Table \ref{tab:sym}.

\begin{table}
    \centering
    \caption{\centering{Mathematical notations used in EnFed}}
    \begin{tabular}{c|c}
    \hline
       Parameters  &  Definition\\
       \hline
       $\beta$ & Size of the request message\\
       $\rho$ & Data transmission rate\\
       $\mathcal{D}_A$  & Local dataset of a device $M$ for application $A$\\
       $\mathcal{A}_A$ & Desired Accuracy Level for application $A$\\ 
      $\mathcal{B}_A$ & Batch size to split the dataset $\mathcal{D}_A$ \\
      $\mathcal{R}_A$ & Maximum number of rounds for updating\\
      & model for application $A$\\ 
      $\mathcal{E}$ & Number of epochs\\
      $N_d$ & Number of nearby devices\\
      $N_{max}$ & Maximum number of contributors \\
      & where $N_{max} \leq N_d$\\
      $N_c$ & Number of contributors\\
      & where $N_c \leq N_{max}$\\
      $B_{min_A}$ & Minimum battery level for training\\
      & for application $A$\\
     $Model_{fin_A}$ & Final Model for device $M$ \\
     & for application $A$\\
     $w$ & Model weights\\
     $E({M_A})$ & Energy consumption of device $M$ \\
     & for updating model for application $A$\\
     $T({M_A})$ & Training time for device $M$ \\
     & for updating model for application $A$\\
     $Accuracy_{M_A}$ & Accuracy of the model\\
     & achieved by device $M$ for application $A$\\
     $B_p$ & Present battery level of the device\\
     $B_{min_A}$ & Minimum battery level to continue model update\\
     & for application $A$\\
     $\mathcal{L}$ & Loss function\\
     \hline
    \end{tabular}
    \label{tab:sym}
\end{table}

\textit{Problem statement:}
Let a device $M$ needs to initialize or update the $model$ of an application $A$, but it does not have sufficient network connectivity and resources to participate in a CFL process, and $M$ does not have sufficient battery level to undergo a DFL process by forming a collaborative network with its nearby devices and exchanging model updates to build a generalized model. $M$ has to create a personalized model that meets the desired accuracy level but at minimal training time and minimal energy consumption, based on the cooperation from nearby devices with updated models. The objective function of the proposed problem is to minimize the training time and the energy consumption of $M$ to update the model for application $A$, and achieve the desired accuracy level, which are represented by equations (\ref{e1}), (\ref{e2}), and (\ref{e3}), as follows:
\begin{equation}
    T_{M_{A}}=min(T(M_A))
    \label{e1}
\end{equation}
where $T_{M_{A}}$ is the minimum training time of $M$ to update the model for application $A$. The training time complexity is illustrated in Section \ref{sectime}.
\begin{equation}
    E_{M_{A}}=min(E(M_A))
    \label{e2}
\end{equation}
where $E_{M_{A}}$ is the minimum energy consumption of $M$ to update the model for application $A$. The energy consumption calculation is illustrated in Section \ref{secen}.
\begin{equation}
    Accuracy_{M_A} \geq \mathcal{A}_A
    \label{e3}
\end{equation}
where $Accuracy_{M_{A}}$ is the prediction accuracy of the model achieved by $M$ for application $A$. The accuracy of a model is defined as $((TP+TN)/(TP+FP+TN+FN))$, where $TP$, $TN$, $FP$, $FN$, denotes true positives, true negatives, false positives, and false negatives, respectively.
\par
\begin{figure*}
\centering
\includegraphics[width=0.8\linewidth, height=2.0in]{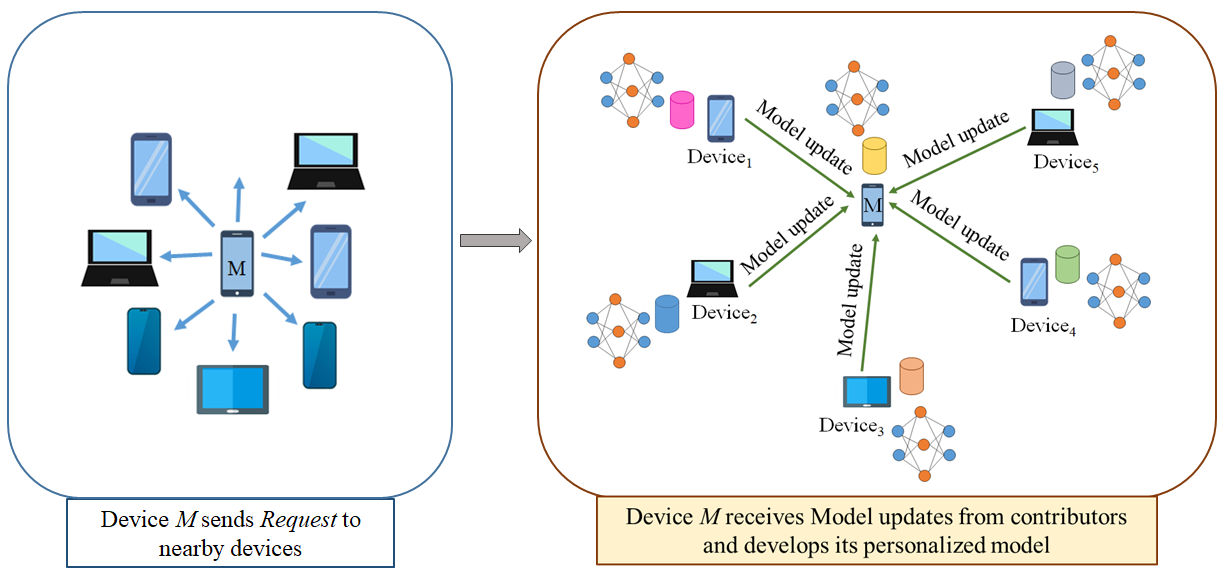}
    \caption{\centering{Requesting nearby devices and receiving model updates from the contributors in EnFed}}
    \label{fl}
\end{figure*}

\textit{Proposed framework:} To solve the stated problem, we use an incentive-based handshaking mechanism to find nearby devices that will help $M$ by sharing their model updates to create its personalized model. The steps of the proposed approach are stated as follows: 
\begin{itemize}
\item \textcolor{black}{$M$ sends a request to its nearby devices ($N_d$) for the $model\;update$, i.e., updated model parameters of the application $A$ with respect to an $incentive$. The type of incentive will depend on the participant. The incentive can be monetary rewards or non-monetary such as, providing tokens for accessing any service. Different types of incentive-based FL schemes exist based on game theory, auction, contract and matching theory, etc. \cite{tu2022incentive}. In our work, we have considered contract theory-based incentive mechanism \cite{tu2022incentive}. The requesting device generates a contract with the contributors, i.e., the nearby devices, which agree with the $incentive$ to share their model updates. During handshaking, the contributors share their keys used for encrypting model updates with $M$, because in EnFed we use the Advanced Encryption Standard (AES)-128, which is a symmetric cryptography approach.}  
\item The devices selected based on the contract theory, send their model updates to $M$. This is assumed that each of the contributing devices has an updated model (using CFL/DFL) for the application. During the transmission of model updates, the model weights are encrypted using AES-128 to prevent leakage of model updates. We use AES-128 because it is a faster encryption algorithm with a lower processing load.
\item The model parameters $M$ received first is used to create the $initial\;model$ after decryption. 
\item $M$ continues to receive model updates from the contributors until the battery level ($B_p$) is below the threshold ($B_{min_A}$) or the number of contributors ($N_c$) reaches its maximum value ($N_{max}$). 
\item After receiving model updates from the contributors, $M$ decrypts them and performs aggregation to build a generalized model. 
\item After aggregation, $M$ fits the model with its own dataset to build its personalized local model and calculates the accuracy score. 
\item If the accuracy score is equal to or greater than the desired accuracy score ($\mathcal{A}_A$), $M$ closes the connections and sets the updated model as the final model. Otherwise, $M$ continues to receive model updates from the contributors, and performs aggregation to update its model until $\mathcal{A}_A$ is reached or $B_p < B_{min_A}$ or the maximum number of rounds ($\mathcal{R}_A$) is reached. 
\end{itemize}
\begin{algorithm}
\small
\caption{Proposed FL algorithm for model update} 
\label{algo_1}
\KwIn{$\mathcal{D}_A$, $\mathcal{B}_A$, $\mathcal{A}_A$, $N_{max}$, $B_{min_A}$, $\mathcal{E}$, $\mathcal{R}_A$}
\KwOut{$Model_{fin_A}$}
\BlankLine
\SetKwFunction{FAlgoName}{Algorithm EnFed}
\SetKwProg{An}{}{}{}
\An{\FAlgoName$()$:} 
{
\If{($N_d \geq 1$)}{$N_c, addr, Keys \gets handshaking()$\\
$modelTrain(N_c, addr, Keys)$
}
}
\An{handshaking():}
{
$sendrequesttonearbydevices(incentive)$\\
    \text{initialize $Keys=[]$, $addr=[]$}, $j \gets 0$\\
    \While{($j < N_{max}$)}
    {
        \If{($agreetocooperate()==TRUE$)}{
            $j \gets j+1$\\
            $acceptconnection(j)$\\
            $createcontract(j)$\\
            $addr.append(connection_j)$\\
            $Keys.append(key_j \gets receivekey(j))$\\
        }
    }
    $N_c \gets j$\\
    $return \; N_c, addr, Keys$
}
\An{modelTrain($N_c, addr, Keys$):} 
{
\text{initialize $modelupdate_c=[]$, $\mathcal{R} \gets 0$}\\
$\mathcal{D}_{A_{train}}, \mathcal{D}_{A_{test}} \gets split(\mathcal{D}_A)$\\
 \For{$j=1 \; to \; N_c$}
  {
   \text{collect $modelupdate_j$}\\
   \text{$modelupdate_j \gets decrypt(modelupdate_j, key_j)$}\\
   \If{($j==1$)}{$model \gets initialize(modelupdate_j)$}
   {$modelupdate_c.append(modelupdate_j)$}\\
   \text{$checkbatterylevel(model, addr, modelupdate_c, j)$}}
   $globalmodel, localmodel \gets updateModel(model, addr, modelupdate_c, N_c)$\\
   $Accuracy \gets accuracy\_score(localmodel, \mathcal{D}_{A_{test}})$\\
   $\mathcal{R} \gets \mathcal{R}+1$\\
   \While{($Accuracy < \mathcal{A}_A$)}
   {
   \text{initialize $modelupdate_u=[]$}\\
    \For{$j=1 \; to\; N_c$}{
    \text{collect $modelupdate_j$}\\
    \text{$modelupdate_j \gets decrypt(modelupdate_j, key_j)$}\\
    $modelupdate_u.append(modelupdate_j)$
   }
   $globalmodel, localmodel \gets updateModel(model, addr, modelupdate_u, N_c)$\\
   $Accuracy \gets accuracy\_score(localmodel, \mathcal{D}_{A_{test}})$\\
   $\mathcal{R} \gets \mathcal{R}+1$\\
   \If{($\mathcal{R}==\mathcal{R}_A$)}{$Model_{fin_A} \gets localmodel$\\
$closeallconnections()$\\
$exit()$}
}
$Model_{fin_A} \gets localmodel$\\
$closeallconnections()$}
\An{checkbatterylevel($model, addr, model_c, j$):}
{
\If{($B_p < B_{min_A}$)}
{$globalmodel, Model_{fin_A} \gets updateModel(model, addr, modelupdate_c, j)$\\
$closeallconnections()$\\$exit()$\\}
}
\An{updateModel($model, addr, model_c, N_c$):}
{
   $m_{u} \gets \sum_{j=1}^{N_c} weights(modelupdate_c[j])/N_c$\\
   $model.setweights(m_u)$\\
   $globalmodel \gets model$\\
   $localmodel \gets model.fit(\mathcal{D}_{A_{train}}, \mathcal{E}, \mathcal{B}_A)$\\
   $return \; globalmodel, localmodel$
}
\end{algorithm}
The proposed FL-based model update method is stated in Algorithm \ref{algo_1}. The requesting device performs handshaking with the nearby devices which agree to share model updates with respect to an incentive. The devices which agree to support are referred to as the contributors. The contributors share their keys used for encrypting model updates with the requesting device. After handshaking, the contributors send their updated model parameters to the requesting device. The device initializes its model based on the model parameters received from the first contributor. The device continues to receive model updates from the contributors until the battery level is below the threshold value or the maximum number of contributors is reached. To check the battery level, \textit{checkbatterylevel} function of Algorithm 1 is used. After receiving model updates from the contributors, it performs aggregation using \textit{updateModel} function of Algorithm 1, and then fits to its training dataset to build its personalized local model. Fig. \ref{fl} shows the pictorial representation of the proposed approach. In the figure, $M$ needs to update its model and $M$ has seven nearby devices. $M$ sends requests to these seven devices. Among them, five devices accept the request and send their model updates to $M$. After receiving model updates $M$ develops its model through aggregation and fitting with its own dataset. This step is repeated until the desired accuracy level is reached, or the battery level drops below the threshold, or the maximum number of rounds is reached. As mobile devices are considered, we assume that each device has at least one device available nearby that can help the device to build its model. Though we have considered mobile devices in this work, IoT devices can also participate in the FL process.   
\subsection{Time complexity}
\label{sectime}
\textcolor{black}{The time complexity and computational complexity of sending requests and receiving responses (agree to cooperate or not) to and from the nearby devices are $O(\beta/\rho)$ and $O(N_d \cdot (\beta/\rho))$, respectively. The time complexity and computational complexity of handshaking with contributors are $O(N_{c})$ and $O(N_{c})$, respectively. The time complexity and computational complexity of sharing the keys are $O(size_{key}/\rho)$ and $O(N_c \cdot (size_{key}/\rho))$, respectively, where $size_{key}$ denotes the size of the key used for the encryption and decryption of model updates.} The time complexity of model initialization is expressed as $O(1)$. The space complexity of model initialization is $O(w)$, where $w$ denotes the model weights. The computational complexity of model initialization is $O(w)$. The time complexity of receiving model updates from contributors is $O(\mathcal{R} \cdot w_j)$, where $w_j$ denotes the model weights of contributor $j$. The computational complexity of receiving model updates from contributors is $O(\mathcal{R} \cdot N_c \cdot w_j)$. \textcolor{black}{The time complexity and computational complexity of encrypting model weights are $O(\mathcal{R} \cdot w_j)$ and $O(\mathcal{R} \cdot N_c \cdot w_j)$, respectively. The time complexity and computational complexity of decrypting model weights are $O(\mathcal{R} \cdot w_j)$ and $O(\mathcal{R} \cdot N_c \cdot w_j)$, respectively.} The time complexity of aggregation is given as $O(\mathcal{R} \cdot N_c \cdot w_j)$. The space complexity of aggregation is $O(N_c \cdot w_j)$. The computational complexity of aggregation is given as $O(\mathcal{R} \cdot N_c \cdot w_j)$. The time complexity of local training is given as $O(\mathcal{R} \cdot \mathcal{E} \cdot (|\mathcal{D}_A|/\mathcal{B}_A) \cdot w)$. The space complexity of local training is $O((|\mathcal{D}_A|/\mathcal{B}_A) \cdot w)$. The computational complexity of local training is given as $O(\mathcal{R} \cdot \mathcal{E} \cdot (|\mathcal{D}_A|/\mathcal{B}_A) \cdot w)$. \\
\textcolor{black}{The total time consumption for training is determined as:}
\begin{equation}
\begin{split}
    T_{train}=T_{dev} + T_{hand} + T_{key} + T_{init} + T_{com} + \\
    T_{enc} + T_{dec} + T_{agg} + T_{loc} 
    \end{split}
    \label{timeequation}
\end{equation}
where $T_{dev}=O(\beta/\rho)$, $T_{hand}=O(N_{c})$, $T_{key}=O(size_{key}/\rho)$, $T_{init}=O(1)$, $T_{com}=O(\mathcal{R} \cdot w_j)$, \textcolor{black}{$T_{enc}=O(\mathcal{R} \cdot w_j)$, $T_{dec}=O(\mathcal{R} \cdot w_j)$}, $T_{agg}=O(\mathcal{R} \cdot N_c \cdot w)$, and $T_{loc}=O(\mathcal{R} \cdot \mathcal{E} \cdot (|\mathcal{D}_A|/\mathcal{B}_A) \cdot w)$.  \par
To meet the objective function presented in equation (\ref{e1}), the minimum value of $\mathcal{R}$, $\mathcal{E}$, and $N_c$, and the maximum value of $\mathcal{B}_A$, which all together meet the condition $Accuracy_{M_A} \geq \mathcal{A}_A$, should be chosen. 

\subsection{Energy consumption of user device}
\label{secen}
\textcolor{black}{The energy consumption is given as the sum of the energy consumption of the mobile device for computation and communication, given as,}
\begin{equation}
    \label{energyequation}
    E_{tot} = E_{comp} + E_{comm}
\end{equation}
where $E_{comp}$ and $E_{comm}$ denotes the energy consumption of the requesting device for computation and communication, respectively.\par
\textcolor{black}{The energy consumption for computation includes energy consumption of the device during model initialization, local model training, aggregation of model updates, and encryption and decryption of model updates. The communication energy consumption contains energy consumption of the requesting device for sending requests to nearby devices, handshaking with the contributors which agree to support, receiving encryption keys from the contributors, and receiving model updates from them. 
The computational energy consumption of the device is given as,}
\begin{equation}
\begin{split}
    E_{comp}=(T_{init} \cdot E_{c_i}) + ((T_{enc} + T_{dec}) \cdot E_c) + (T_{agg} \cdot E_{c_a}) + \\
    (T_{loc} \cdot E_{c_l})
\end{split} 
\end{equation}
where $E_{c_i}$, $E_c$, $E_{c_a}$, and $E_{c_l}$ presents the average power consumption of the device per unit time during model initialization, encryption and decryption of model updates, aggregation of model updates, and local model training, respectively.
\textcolor{black}{The communication energy consumption of the device is given as,}
\begin{equation}
    E_{comm}=((T_{dev} + T_{hand}) \cdot E_s) + ((T_{hand} + T_{key} + T_{com}) \cdot E_r)
\end{equation}
where $E_s$ and $E_r$ presents the average power consumption of the device per unit time during transmission and reception modes, respectively. Here, the energy consumption during the periods of finding nearby devices, handshaking, receiving encryption keys, and receiving model updates, are considered. 
\par
\textcolor{black}{To meet the objective function presented in equation (\ref{e2}), the following condition is to be satisfied:}\\ 
\begin{equation}
E_{tot} = E_{M_A}
\end{equation}
\textcolor{black}{This condition will meet if the $T = T_{M_A}$, where} 
\begin{equation}
    T = T_{train}
\end{equation}
\textcolor{black}{Here, the training time $T_{train}$ includes the time consumptions in finding nearby devices, handshaking, sharing encryption keys, model initialization, encryption and decryption of model updates during transmission, receiving updates from contributors, aggregation, and fitting with own dataset.} As the energy consumption of the device at transmission, reception, and execution modes are different, and the values vary depending on the device's state, configuration, number of present computations, etc., the objective should be to minimize the time consumption, so that the energy consumption can be minimized. 

\subsection{Convergence in EnFed}
\textcolor{black}{The objective of the learning approach is to minimize the loss functions $\mathcal{L}_1(w_{M_A})$ and $\mathcal{L}_2(w_{M_A})$.
$\mathcal{L}_1(w_{M_A})$ denotes the loss function for the aggregated model updates and $\mathcal{L}_2(w_{M_A})$ denotes the loss function for the local model update using own dataset. However, to further discuss the convergence of EnFed, firstly, we have to show that each contributor $j$ has a local model with minimal loss.} 
\par 
\textcolor{black}{\textit{Local model at contributor $j$:} A contributor $j$ sends its local model update to the requesting device if it has a local model with minimal loss. If $w_{j_A}^{q}$ denotes the model weights for application $A$ at epoch $q$, then}
\begin{align}
w_{j_A}^{q+1} = w_{j_A}^{q} - \eta_j\nabla w_{j_A}^{q}
\end{align}
\textcolor{black}{where $1 \leq q \leq \mathcal{E}_j$, $\eta_j$ is the learning rate, and $\mathcal{E}_j$ denotes the number of local epochs for a contributor $j$. The loss function $\mathcal{L}(w_{j_A})$ is given as:}
\begin{align}
\mathcal{L}(w_{j_A})=\mathcal{L}(w_{j_A}^{q}) - \mathcal{L}(w_{j_A}^{q+1})
\end{align}
\textcolor{black}{$\mathcal{L}(w_{j_A})$ will be minimum if the following condition is satisfied:} 
\begin{align}
(\mathcal{L}(w_{j_A}^{q}) - \mathcal{L}(w_{j_A}^{q+1})) \to 0
\end{align}
\textcolor{black}{The above condition will be satisfied when $q=\mathcal{E}_j$. Therefore,}  
\begin{align}
\label{locloss}
   (\mathcal{L}(w_{j_A}^{\mathcal{E}_j}) - \mathcal{L}(w_{j_A}^{\mathcal{E}_j+1})) \to 0
\end{align}
\textcolor{black}{Therefore, the loss function $\mathcal{L}(w_{j_A})$ is minimum when $q=\mathcal{E}_j$, given as, $\mathcal{L}(w_{j_A}) \to 0$ at $q=\mathcal{E}_j$. At a round $r$ the contributor $j$ sends its local model update with minimal loss to the requesting device.} 

\textcolor{black}{\textit{Aggregation:} The requesting node aggregates the local model updates received from the contributors at each round. The aggregated model update at a round $r$ is given as,}
\begin{align}
w_{M_A}^{r+1} = \frac{1}{N_c} \sum_{j=1}^{N_c}{w_{j_A}^{r}}
\end{align}

\textcolor{black}{The loss function for the aggregated model updates, $\mathcal{L}_1(w_{M_A})$, is given as:}
\begin{align}
\mathcal{L}_1(w_{M_A}) = \frac{1}{N_c} \sum_{j=1}^{N_c}\mathcal{L}(w_{j_A})
\end{align}
\textcolor{black}{where $w_{M_A}$ denotes the aggregated model weights of the device $M$ for the application $A$, and $N_c$ is the number of contributors.}\\
\textcolor{black}{From equation (\ref{locloss}), we observe that the contributor $j$ has its local model with minimal loss when $q=\mathcal{E}_j$. If a contributor $j$ has a local model with minimal loss, then it sends its model update to $M$. Hence, for the contributor $j$ the following equation is satisfied \cite{mukherjee2024federated}:}
\begin{align}
\lim_{\mathcal{R} \to \infty} \frac{1}{\mathcal{R}} \sum_{r=1}^{\mathcal{R}} \mathbb{E}[\|\nabla \mathcal{L}(w_{j_A}^r)\|^2] = 0
\end{align}
\textcolor{black}{where $\mathcal{R}$ denotes the number of rounds.} \par 
\textcolor{black}{As each contributor has its model with minimal loss, thus, $(\mathcal{L}(w_{j_A}^{\mathcal{R}-1}) - \mathcal{L}(w_{j_A}^{\mathcal{R}}))$ approaches zero.
This demonstrates that the local model of $j$ meets the convergence point. Hence, the individual loss of each contributor is minimal, i.e. $\mathcal{L}(w_{j_A})=\mathcal{L}_{min}(w_{j_A})$. 
Therefore,}  
\begin{align}
\mathcal{L}_1(w_{M_A}) = \frac{1}{N_c} \sum_{j=1}^{N_c}\mathcal{L}_{min}(w_{j_A})
\end{align}
\textcolor{black}{Hence, $\mathcal{L}_1(w_{M_A}) = \mathcal{L}_{min}(w_{M_A})$.}
\par
\textcolor{black}{\textit{Local model at the requesting device:} After aggregation, the requesting device fits the aggregated model with own dataset. If $w_{M_A}^{q}$ denotes the model weights for application $A$ at epoch $q$, then}
\begin{align}
w_{M_A}^{q+1} = w_{M_A}^{q} - \eta\nabla w_{M_A}^{q}
\end{align}
\textcolor{black}{where $1 \leq q \leq \mathcal{E}$ and $\eta$ is the learning rate. The loss function $\mathcal{L}_2(w_{M_A})$ is given as:}
\begin{align}
\mathcal{L}_2(w_{M_A})=\mathcal{L}(w_{M_A}^{q}) - \mathcal{L}(w_{M_A}^{q+1})
\end{align}
\textcolor{black}{$\mathcal{L}_2(w_{M_A})$ will be minimum if the following condition is satisfied:} 
\begin{align}
(\mathcal{L}(w_{M_A}^{q}) - \mathcal{L}(w_{M_A}^{q+1})) \to 0
\end{align}
\textcolor{black}{If $q=\mathcal{E}$, then}  
\begin{align}
\label{loclosso}
   (\mathcal{L}(w_{M_A}^{\mathcal{E}}) - \mathcal{L}(w_{M_A}^{\mathcal{E}+1})) \to 0
\end{align}
\textcolor{black}{Therefore, the loss function $\mathcal{L}_2(w_{M_A})$ will be minimum when $q=\mathcal{E}$, given as, $\mathcal{L}_2(w_{M_A}) \to 0$ at $q=\mathcal{E}$. Hence, after reaching the number of epochs, the local model of the device meets the convergence point.}

\subsection{\textcolor{black}{Difference of EnFed from Opportunistic FL, Participatory FL, and Asynchronous FL}}
\textcolor{black}{In opportunistic federated learning (OFL) \cite{lee2021opportunistic}, (i) the devices continuously look for nearby devices to obtain local model updates of the neighbours, (ii) then use the received model updates to continuously update its own model using local dataset. The difference of EnFed from the OFL is that no incentive is provided to the neighbours in OFL, whereas in EnFed the supporting devices having trained models are offered incentive to share their model updates with the requesting device, so that it can build its local model based on the received updates from the supporting devices. In participatory federated learning (PFL) \cite{buratto2024energy}, a central server selects a subset of clients per FL round, which send their local model updates to the central server that aggregates the received updates to build a global model. In asynchronous federated learning (AFL) \cite{xu2023asynchronous}, the devices with heterogeneous computational ability train their local models and send their model updates to the central server independently without waiting for others. The server performs buffered aggregation or weighted averaging to build the global model.  The difference of EnFed from the PFL and AFL is that (i) in EnFed no central server performs aggregation to build a global model, (ii) each supporting node has a trained model and it sends the local model updates to a resource-limited node that requests for support to build its model with respect to an incentive. Hence, we observe that EnFed is different from OFL, PFL, and AFL, and unique, where (i) a resource-limited device with limited battery life requests nearby devices to send their model updates with respect to an incentive, (ii) the devices which agree to the offered incentive help the device by sending their model updates, (iii) the requesting device builds its local model by aggregating the received updates, and then fitting with its own dataset.} 

\section{Performance Evaluation}
\label{perf}
\begin{figure}
\centering
    \includegraphics[width=0.99\linewidth, height=2.8in]{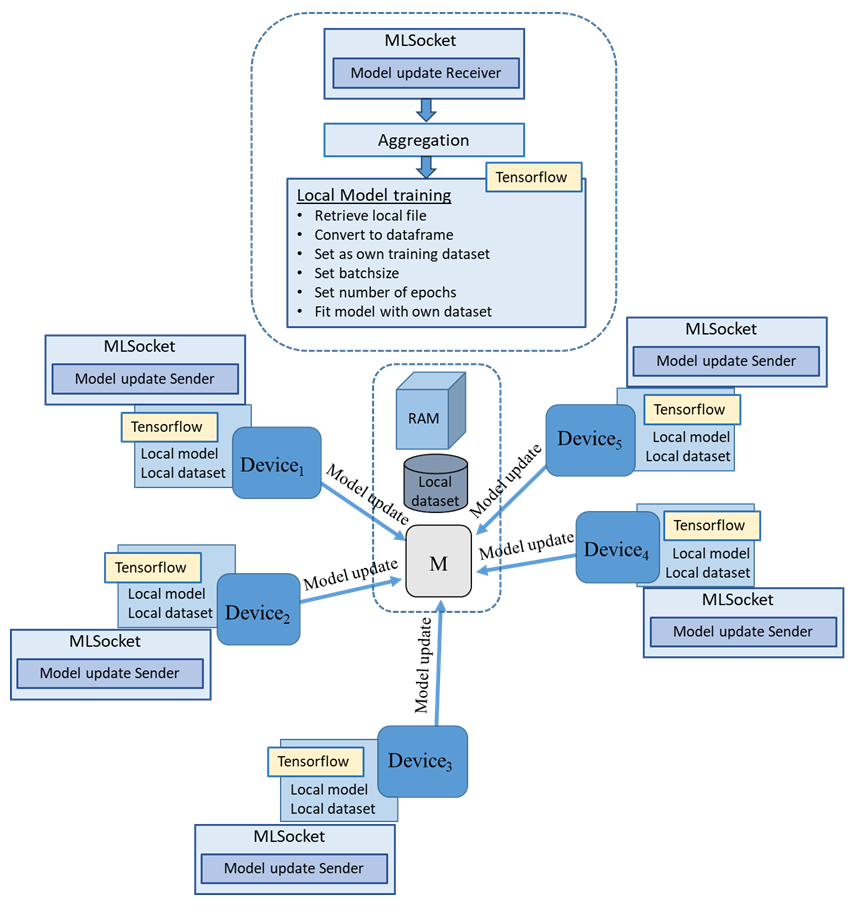}    \caption{\centering{Implementation diagram of EnFed}}
    \label{actimplement}
\end{figure}

To implement EnFed, Python 3.8.10 is used, and Tensorflow is used for DL models. To connect the devices for transmitting and receiving model updates, \textit{MLSocket} is used. The implementation diagram is presented in Fig. \ref{actimplement}. The experimental setup is explained in Section \ref{exs}. In EnFed, the supporting devices have their local models and local datasets. The supporting devices send their model updates to the requesting device. The requesting device receives model updates from them, and develops its personalized model. \textcolor{black}{For local data analysis, LSTM and MLP are used in EnFed. GRU is also popular to model and process sequential data. However, we prefer LSTM over GRU because (i) LSTM features a separate cell state that enables to keep significant information over longer time period, (ii) The use of three gates (input, forget, and output) in LSTM provides a fine-grained control over information flow that helps to outperform GRU on large-scale and complex problems, (iii) To deal with intricate and long-term dependencies within sequences, LSTM provides better performance especially for deep time-series forecasting. CNN is also a popular data analysis algorithm especially for activity recognition. However, instead of using CNN, we consider LSTM because (i) To capture long-term dependencies with sequential data, LSTM performs better, (ii) LSTM excels in processing time series data by maintaining a memory of past inputs, (iii) LSTMs can handle input sequences of varying lengths, (iv) By using three gates, LSTM can focus on the most significant parts of a sequence, (v) LSTM can better understand the overall context of a sequence. MLP is also used in EnFed for local data analysis because MLP offers benefits with structured data, lower computational resources, reduced overfitting on small dataset, and can be faster due to its reduced overhead associated with complex convolutional layers.}
The values of the parameters used in classification using LSTM and MLP are presented in Table \ref{tab:class}. We determine the prediction accuracy for the proposed approach EnFed along with training time and energy consumption. After that we compare the performance of EnFed with conventional CFL and DFL-based systems (as implemented in \cite{mukherjee2024federated}) with respect to training time and energy consumption to achieve the desired accuracy level. \textcolor{black}{For aggregation, federated averaging (FedAvg) is used in the implementation of CFL and DFL.} Finally, the accuracy and response time in EnFed are compared to the conventional cloud-only framework.   

\begin{table}[]
\centering
\caption{\centering{The values of the parameters for model classification}}
\begin{tabular}{lll}
\hline
Classifier & Parameter & Value \\
\hline
LSTM       & \begin{tabular}[c]{@{}l@{}}Activation:\\Optimizer:\\Loss:\\Epochs:\\\end{tabular}                    & \begin{tabular}[c]{@{}l@{}}softmax\\Adam\\Categorical crossentropy\\100\end{tabular}   \\
\hline
MLP       & \begin{tabular}[c]{@{}l@{}}Hidden layer sizes:\\Activation:\\ Maximum iteration:\\Solver:\\\end{tabular}                    & \begin{tabular}[c]{@{}l@{}}(64, 32)\\ReLU\\100\\Adam\end{tabular}   \\
\hline
\end{tabular}
\label{tab:class}
\end{table}

\subsection{Experimental setup}
\label{exs}
We have carried out experiments using resources accessible from the Melbourne CLOUDS Lab. We have provisioned eight (8) virtual machines (VMs) created over the Australia RONIN cloud environment, which is hosted on Amazon AWS. Among them, one is used as the requesting node, and five (5) VMs are utilized as the supporting nodes. Each of the nodes has 4GB memory. We have used the remaining two VMs each having 4GB memory and 2vCPUs, as follows: one is used as the edge server, and the other one is used as the cloud server instance. For performance analysis, we have used two different datasets. The first dataset\footnote{\url{https://www.kaggle.com/datasets/aadhavvignesh/calories-burned-during-exercise-and-activities}} contains different types of physical activities and the respective calories burned. In our experiment, we categorize calories according to the range ($<0.5, 0.5-1, 1.0-2.0, 2.0-3.0, >3.0$) for classification. \textcolor{black}{The $dataset^1$ is non-identically distributed among the requesting node and five supporting nodes.} The second dataset\footnote{\url{https://www.kaggle.com/datasets/nurulaminchoudhury/harsense-datatset}} is a HAR dataset that contains the movement data of twelve users. The movement data was collected using accelerometer and gyroscope. Based on the movement data, the activity of the user is predicted. For the second dataset, six types of activities are considered: Running (1), Walking (2), Sitting (3), Standing (4), Downstairs (5), and Upstairs (6). \textcolor{black}{The $dataset^2$ is non-identically distributed among the requesting node and five supporting nodes.} The training time is measured in seconds (s) and the energy consumption of the device during the period is measured in Joule (J). The training time and energy consumption of the device are measured using equations (\ref{timeequation}) and (\ref{energyequation}), respectively. For securing the model updates during transmission, we have used AES-128-based encryption. In the designed framework, for communication among the devices, Orthogonal Frequency Division Multiple Access (OFDMA) is used.  

\subsection{Performance of EnFed using LSTM}
The desired accuracy level for the experiment is set to 0.95 and the battery threshold is considered 20\%. As we have five VMs in the experiment as the supporting nodes, the maximum number of contributors is set to five. In this regard, we would like to mention that as the requesting node is a mobile device, the number of nearby devices will not be high. Furthermore, the objective is to build a model with the desired accuracy level at a minimal training time and minimal energy consumption of the device, therefore, the maximum number of contributors will also not be high.\par

\begin{figure}
\centering
\includegraphics[width=0.6\linewidth, height=1.5in]{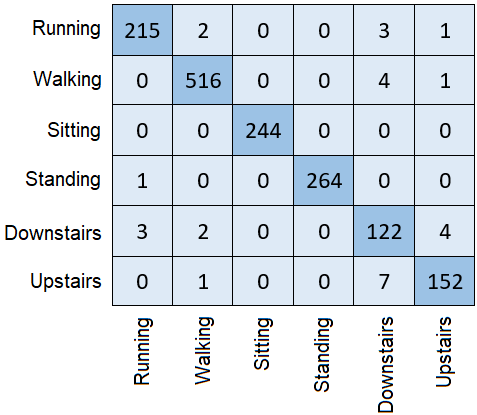}
    \caption{\centering{Confusion matrix of EnFed for Second Dataset (Number of contributors: 05)}}
    \label{conf}
\end{figure}

\begin{figure}
\centering
    \includegraphics[width=0.99\linewidth, height=2.0in]{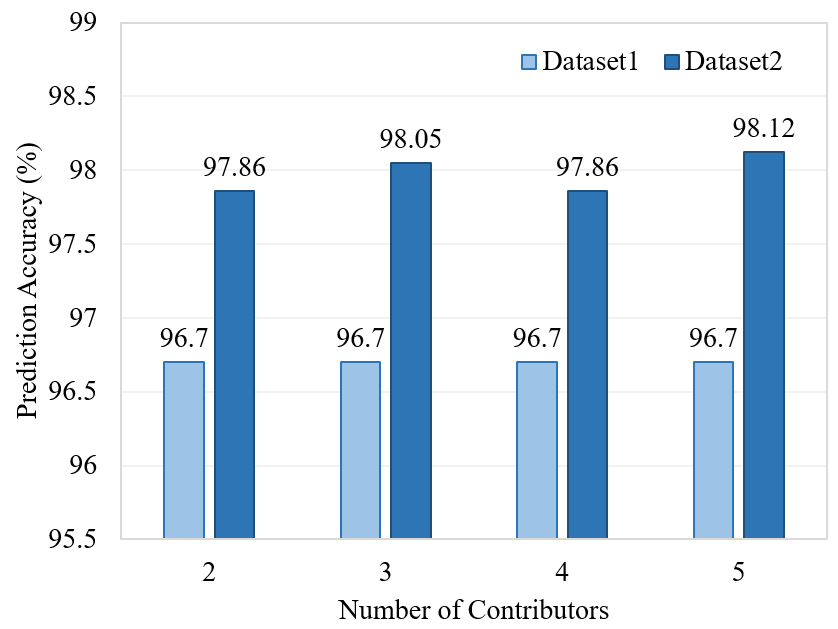}
    \caption{\centering{Prediction accuracy in EnFed}}
    \label{acc}
\end{figure}

\begin{figure}
\centering
    \includegraphics[width=0.99\linewidth, height=2.0in]{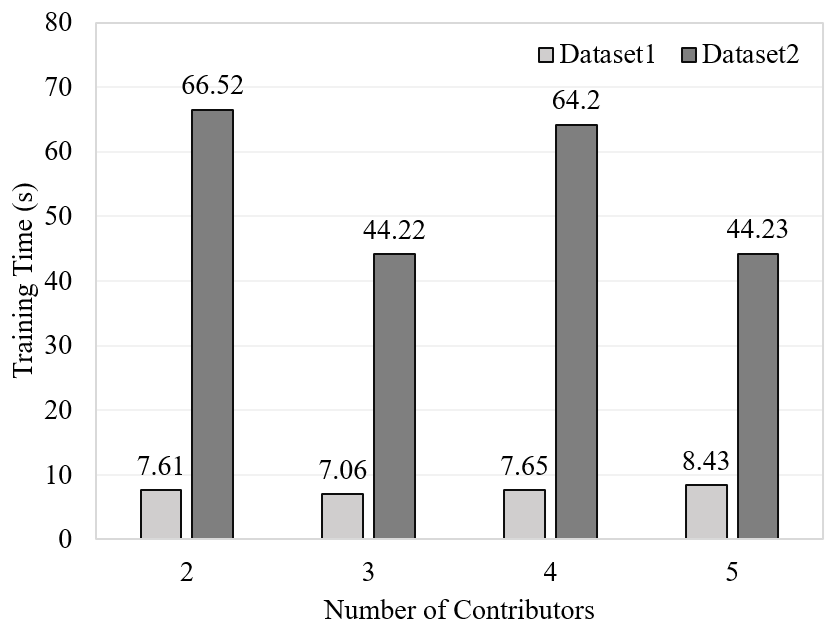}
    \caption{\centering{Training time in EnFed}}
    \label{traintime}
\end{figure}

\begin{figure}
\centering
    \includegraphics[width=0.99\linewidth, height=2.0in]{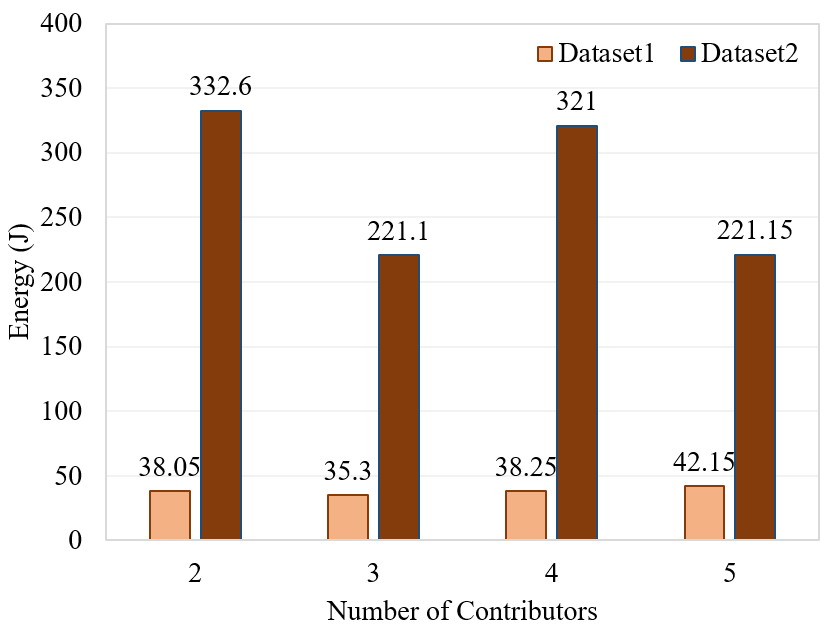}
    \caption{\centering{Energy consumption of the device in EnFed}}
    \label{trainen}
\end{figure}

For the first dataset, EnFed has achieved a prediction accuracy of 96.7\% for two, three, four, and five contributors, and the precision, recall, and F1-score are above 0.9. The training time for two, three, four, and five contributors are 7.61s, 7.06s, 7.65s, and 8.43s, respectively. Thus, the average training time is 7.69s. The energy consumption of the device for two, three, four and five contributors are 38.05J, 35.3J, 38.25J, and 42.15J, respectively. The average energy consumption of the device is 38.44J. For the second dataset, EnFed has achieved 97.86\%, 98.05\%, 97.86\%, and 98.12\% accuracy for two, three, four, and five contributors, respectively. We observe that for five contributors, the highest accuracy of 98.12\% (0.98) is achieved. We also observe that the precision, recall, and F1-Score are 0.967. The respective confusion matrix is presented in Fig. \ref{conf}. The training time for two, three, four, and five contributors are 66.52s, 44.22s, 64.2s, and 44.23s, respectively. Thus, the average training time is 54.8s. The energy consumption of the device for two, three, four, and five contributors are 332.6J, 221.1J, 321J, and 221.15J, respectively. Thus, the average energy consumption of the device is 273.96J. The prediction accuracy, training time, and energy consumption in EnFed using LSTM are presented in Figs. \ref{acc}, \ref{traintime}, and \ref{trainen}, respectively. We observe that the average prediction accuracy for the first and second datasets are $\sim$97\% and $\sim$98\% respectively. The precision, recall, and F1-Score, are also above 0.9 for all cases for the first dataset and above 0.95 for all cases 
for the second dataset.


\par
\begin{figure*}
    \centering
    \begin{subfigure}[b]{0.49\textwidth}
        \centering
\includegraphics[width=0.9\linewidth, height=2.0in]{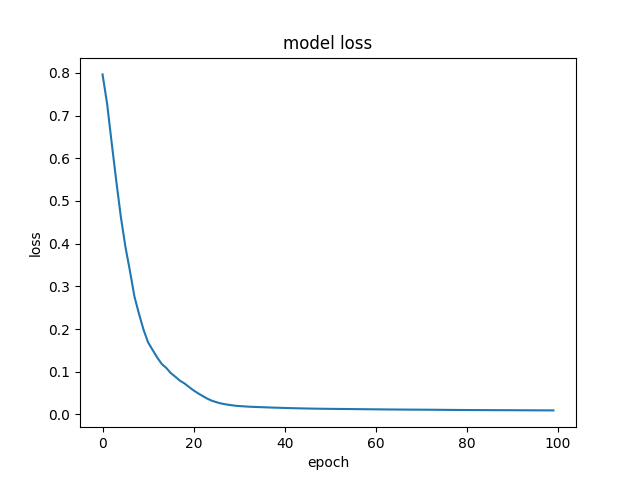}
        \caption{Local model loss for first dataset}
        \label{loss1}
    \end{subfigure}
    \hfill
    \begin{subfigure}[b]{0.49\textwidth}
        \centering
\includegraphics[width=0.9\linewidth, height=1.8in]{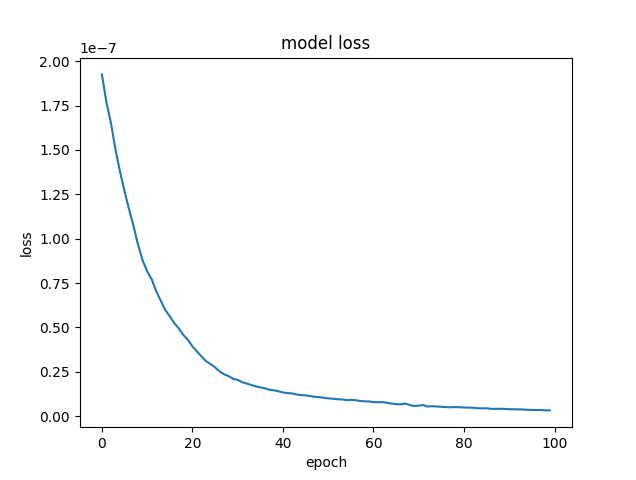}
        \caption{Local model loss for second dataset}
        \label{loss2}
    \end{subfigure}
    \caption{Local model loss in EnFed}
    \label{loss}
\end{figure*}

The local model loss in EnFed after model update using five contributors is presented in Fig. \ref{loss}. In EnFed, the number of rounds is decided based on the desired accuracy level and the battery level of the requesting device. However, a maximum number of rounds (10) is also set to prevent the process from going through an infinite number of iterations. 
In our experiment, the desired accuracy level is achieved in the first, second, or third round, for all the cases. As we observe from the figure, for both the datasets the process converges when the loss is minimal. We have also evaluated the performance of EnFed for another activity recognition dataset\footnote{\url{https://www.kaggle.com/datasets/uciml/human-activity-recognition-with-smartphones}}. This dataset contains sensor values including accelerometer and gyroscope values, based on which the user activity is predicted. The dataset contains data of 30 users, and there are six categories of activities: standing, sitting, laying, walking, walking downstairs, and walking upstairs. For this dataset, EnFed using LSTM has achieved prediction accuracy of $>$98\%, with a precision, recall, and F1-score of above 0.98.

\subsubsection*{Comparison with DFL and CFL}
The performance of EnFed in terms of prediction accuracy, training time, and energy consumption is compared with the conventional DFL and CFL-based systems. In a conventional DFL framework, the devices form a collaborative network and exchange model updates among themselves until a desired generalized model is developed. In a conventional CFL system, the devices exchange their model updates with the server until a desired global model is developed. The comparison of performance of EnFed, conventional DFL, and CFL systems, are presented in Table \ref{tab:com1}. We observe that EnFed has achieved the desired accuracy level at much less training time and energy consumption compared to the conventional DFL and CFL systems. The number of clients is considered six in CFL. The number of nodes in conventional DFL is considered three, four, five, and six. The average prediction accuracy, training time, and energy consumption are determined, and presented in Table \ref{tab:com1}. For DFL, both mesh and ring topologies (peer-to-peer) are considered, and the average values of the obtained results are presented in the table. \par
For the \textit{first dataset}: (i) the average training time for DFL is 18.73s (the average training time for DFL with three, four, five, and six devices are 18.15s (mesh: 18.15s, ring: 18.15s), 13.5s (mesh: 16.2s, ring: 10.8s), 20.52s (mesh: 24.624s, ring: 16.416s), and 22.75s (mesh: 27.3s, ring: 18.2s), respectively), (ii) the accuracy level achieved by DFL in all four cases is 96.7\%, i.e., $\sim$97\%, (iii) the training time for CFL is 50.04s, and it has achieved an accuracy of 99.9\%, (iv) the average energy consumption of the device for DFL is 93.65J (the average energy consumption of the device for DFL with three, four, five, and six devices are 90.75J (mesh: 90.75J, ring: 90.75J), 67.5J (mesh: 81J, ring: 54J), 102.6J (mesh: 123.12J, ring: 82.08J), and 113.75J (mesh: 136.5J, ring: 91J), respectively), and (v) the energy consumption of the device for CFL is 250.2J. \par
For the \textit{second dataset}: (i) the average training time for DFL is 67.74s (the average training time for DFL with three, four, five, and six devices are 70.14s (mesh: 70.14s, ring: 70.14s), 61.8s (mesh: 67.98s, ring: 55.62s), 66.76s (mesh: 68.02s, ring: 65.5s), and 72.26s (mesh: 79.49s, ring: 65.03s), respectively), (ii) the prediction accuracy achieved by DFL is $\sim$98\% (the accuracy level achieved by DFL with three, four, five, and six devices are 97.98\%, 98.12\%, 97.98\%, and 98.31\%, respectively),  (iii) the training time for CFL is 75.33s, and it has achieved an accuracy of 98.39\%, (iv) the average energy consumption of the device for DFL is 338.7J (the average energy consumption of the device for DFL with three, four, five, and six devices are 350.7J (mesh: 350.7J, ring: 350.7J), 309J (mesh: 339.9J, ring: 278.1J), 333.8J (mesh: 340.1J, ring: 327.5J), and 361.3J (mesh: 397.45J, ring: 325.15J), respectively), and (v) the energy consumption of the device for CFL is 376.65J. \par
We observe that CFL has achieved the highest accuracy for both datasets. However, the training time and energy consumption of the device are much high, when CFL is used. In DFL, the training time and energy consumption are lower than CFL. Though we have considered both the ring and mesh topologies, we observe that the training time and energy consumption are much less in ring-based DFL than the mesh-based DFL, due to lower communication overhead. We also observe that in EnFed, the prediction accuracy is almost same as compared to both mesh-based and ring-based DFL frameworks but with lower training time and lower energy consumption. As we observe, EnFed has achieved above 96\% accuracy for both datasets. The average training time in EnFed and DFL are presented in Table \ref{tab:com1}. From the results we observe that using EnFed the training time and energy consumption are reduced by approximately 59\% and 19\% than DFL for the first and second datasets, respectively. We also observe that EnFed has reduced the training time and energy consumption by approximately 85\% and 27\% than CFL for the first and second datasets, respectively. 
\begin{table} []
\centering
\caption{\centering{Comparison of accuracy, training time, and energy consumption in EnFed, DFL, and CFL, using LSTM}}
\label{tab:com1}
\begin{tabular}{ |c|c|c|c| } 
\hline
Performance Metric & EnFed & DFL & CFL\\
\hline
\multirow{3}{4em}{Accuracy} & $\sim$97\% & $\sim$97\% & 99.9\%\\ 
& (Dataset1) & (Dataset1) & (Dataset1)\\ 
& $\sim$98\% & $\sim$98\% & 98.39\%\\
& (Dataset2) & (Dataset2) & (Dataset2)\\ 
\hline
\multirow{4}{4em}{Training time} & 7.69s & 18.73s & 50.04s\\ 
& (Dataset1) & (Dataset1) & (Dataset1)\\ 
& 54.8s & 67.74s & 75.33s\\
& (Dataset2) & (Dataset2) & (Dataset2)\\
& (Average) & (Average) & \\ 
\hline
\multirow{4}{4em}{Energy consumption} & 38.44J & 93.65J & 250.2J\\ 
& (Dataset1) & (Dataset1) & (Dataset1)\\ 
&273.96J & 338.7J & 376.65J\\
& (Dataset2) & (Dataset2) & (Dataset2)\\ 
& (Average) & (Average) & \\ 
\hline
\end{tabular}
\end{table}

\subsection{Performance of EnFed using MLP}
For further validation, we consider MLP also as an underlying approach for data analysis, and in this case we consider five contributors. We observe that for MLP, EnFed has achieved: (i) a prediction accuracy of 96\% for the first dataset and 98.05\% for the second dataset, (ii) the training time for the first and second datasets are 4.28s and 4.29s respectively, and (iii) the energy consumption for the first and second datasets are 21.4J and 21.45J respectively. The precision, recall, and F1-Score, are also above 0.9 for the first dataset and above 0.95 for the second dataset. For another activity recognition $dataset^3$, EnFed using MLP has achieved $>$98\% prediction accuracy, with a precision, recall, and F1-score of above 0.98.

\subsubsection*{Comparison with DFL and CFL}
The results of EnFed using MLP are compared with the conventional DFL and CFL systems in Table \ref{tab:com2}. In the case of DFL using MLP, 96\% accuracy is achieved for the first dataset, the average training time is 9.66s (mesh: 11.59s, ring: 7.73s), and the energy consumption of the device is 48.3J (mesh: 57.95J, ring: 38.65J). For the second dataset DFL has achieved 98.5\% accuracy using MLP, the training time is 15.67s (mesh: 18.8s, ring: 12.54s), and the energy consumption of the device is 78.35J (mesh: 94J, ring: 62.7J). In case of CFL using MLP, for the first dataset 99.9\% accuracy is achieved, the training time is 51.5s, and the energy consumption of the device is 257.5J. For the second dataset CFL has achieved 98.5\% accuracy using MLP, the training time is 102.63s, and the energy consumption of the device is 513.15J. 

\begin{table} []
\centering
\caption{\centering{Comparison of accuracy, training time, and energy consumption in EnFed, DFL, and CFL, using MLP}}
\label{tab:com2}
\begin{tabular}{ |c|c|c|c| } 
\hline
Performance Metric & EnFed & DFL & CFL\\
\hline
\multirow{3}{4em}{Accuracy} & 96\% & 96\% & 99.9\%\\ 
& (Dataset1) & (Dataset1) & (Dataset1)\\ 
& 98.05\% & 98.5\% & 98.5\%\\
& (Dataset2) & (Dataset2) & (Dataset2)\\ 
\hline
\multirow{4}{4em}{Training time} & 4.28s & 9.66s & 51.5s\\ 
& (Dataset1) & (Dataset1) & (Dataset1)\\ 
& 4.29s & 15.67s & 102.63s\\
& (Dataset2) & (Dataset2) & (Dataset2)\\ 
\hline
\multirow{4}{4em}{Energy consumption} & 21.4J & 48.3J & 257.5J\\ 
& (Dataset1) & (Dataset1) & (Dataset1)\\ 
&21.45J & 78.35J & 513.15J\\
& (Dataset2) & (Dataset2) & (Dataset2)\\ 
\hline
\end{tabular}
\end{table}

As we observe, for MLP also EnFed has achieved above 95\% accuracy but with lower training time and lower energy consumption than CFL and DFL. Using EnFed, the training time and energy consumption are reduced by approximately 92\% for the first dataset and 96\% for the second dataset, than CFL. Compared to DFL, EnFed has approximately 56\% and 73\% lower training time and lower energy consumption, for the first and second datasets, respectively. 

\subsection{Real Setup and Simulation Results}
\textcolor{black}{To depict a real scenario, we have performed an experimental analysis using eight laptops, each with 8GB RAM and 1TB HDD. Among these laptops, one is used as the requesting device, and it selects a maximum of five contributors among seven other laptops, and builds its model using the proposed approach. We observe that for the first dataset: (i) the average training time is 10.55s, the average energy consumption of the requesting device is 52.75J, and the average prediction accuracy is 97\%, if LSTM is used in EnFed, (ii) the average training time is approximately 4.3s, energy consumption is 22J, and prediction accuracy is 96\%, if MLP is used in EnFed. For the second dataset: (i) the average training time is 65.1s, the average energy consumption of the requesting device is 330J, and the average prediction accuracy is 98\%, if LSTM is used in EnFed, (ii) the average training time is 4.4s, the average energy consumption of the requesting device is 23J, and the average prediction accuracy is 98\%, if MLP is used in EnFed. 
For further validation of EnFed, we have performed a simulation also considering 100 nodes (each with maximum 15 nearby devices and maximum 10 contributors) based on the configuration of a mobile device with an average power consumption of 5 watts per unit time. As the requesting node is a mobile device, the maximum number of nearby nodes is considered 15, and maximum number of supporting nodes is considered 10 in simulation. We observe that for the first dataset: (i) the average training time is approximately 12.5s, energy consumption is 62J, and prediction accuracy is 97\%, if LSTM is used in EnFed, (ii) the average training time is approximately 4.5s, energy consumption is 23J, and prediction accuracy is 96\%, if MLP is used in EnFed. For the second dataset: (i) the average training time is approximately 65s, energy consumption is 330J, and prediction accuracy is 98\%, if LSTM is used in EnFed, (ii) the average training time is approximately 4.5s, energy consumption is 23J, and prediction accuracy is 98\%, if MLP is used in EnFed.}

\subsection{Comparison with Other Classifiers}
\textcolor{black}{We have also evaluated the performance of EnFed with two well-known classifiers GRU and CNN. We observe that using GRU and CNN approximately (i) 75\% and 60\% accuracy are achieved for the first dataset, respectively, and (ii) 80\% and 70\% accuracy are achieved for the second and third datasets, respectively. The achieved accuracy values are quite lower than EnFed using LSTM and MLP. The more complex and specialized structure of LSTM and the use of three gates and a separate cell state in LSTM, has provided better performance compared to GRU and CNN. On the other hand, the ability of MLP to learn complex patterns also helps it to outperform EnFed with GRU and CNN.}

\subsection{Comparison with existing activity recognition models}
In Table \ref{tab:2}, we have compared the proposed approach with the existing activity recognition approaches. 
Most of the existing approaches mainly focused on prediction accuracy rather the training time and energy consumption. However, in our system we have considered these two metrics because for real-time applications time and energy consumption are vital. As we observe from Table \ref{tab:2}, EnFed has achieved a high prediction accuracy for the datasets, which we have considered in the experiment. Furthermore, EnFed has determined the training time and energy consumption, and we observe that the training time and energy consumption of the device during the training period are not very high. Furthermore, user data are not shared and model weights are encrypted during the transmission of model updates to protect data privacy. Thus, we can refer to EnFed as a privacy-aware, energy-efficient, and fast activity monitoring and recognition system.

\begin{table}
\caption{\centering{\textcolor{black}{Comparison of EnFed with existing activity recognition systems}}}
    \centering
    \begin{tabular}{c c c c c}
        \hline
         Work & Use of FL& Accuracy & Training & Energy \\
        & & & time & consumption\\
        \hline
    Tang & No & 79.02\%- & Not & Not\\
      et al.  \cite{tang2022multiscale}& & 99.02\%& determined & determined\\
      Ferrari & No & 79.49\% & Not & Not\\
      et al.  \cite{ferrari2023deep}& & & determined & determined\\
      Priyadarshini & No & 97.998\% & Not & Not\\
      et al.  \cite{priyadarshini2023human}& & & determined & determined\\
      Kaya & No & 90.27\%- & Not & Not\\
      et al.  \cite{kaya2024human}& & 98\% & determined & determined\\
      Jha & No & 87.81\%- & Not & 0.1139 Milli\\
      et al.  \cite{jha2022hybrid}& & 98.57\%& determined & Amp Hour\\
      Zou & No & 82.29\%- & Not & Not\\
      et al.  \cite{zou2024gt}& & 96.71\%& determined & determined\\
      Guo & No & $>$95\% & Not & Not\\
      et al.  \cite{guo2021evolutionary}& & & determined & determined\\
      Cheng & Yes & 87.727\%- & Not & Not\\
      et al.  \cite{cheng2023protohar}& (proto-type& 95.11\%& determined & determined\\
      & based FL)& & & \\
       Shaik  & Yes & 98\%- & Not & Not\\
      et al.  \cite{shaik2022fedstack}& (FedStack) & 99\%& determined & determined\\
      Albomany & Yes & 96\% & Not & Not\\
      \cite{albogamy2025federated} &(FedAvg) & & determined & determined\\
      Proposed & Yes & 96\%- & 4.28s- & 21.4J-\\
      work (EnFed)& & 98.05\% & 54.8s & 273.96J\\
       \hline
    \end{tabular}
    \label{tab:2}
\end{table}

\begin{table}[]
    \centering
    \caption{\centering{Prediction accuracy in EnFed and Cloud-only system without FL}}
    \begin{tabular}{ |c|c|c| } 
\hline
Approach & Prediction accuracy & Prediction accuracy \\
& for Dataset1 & for Dataset2\\
\hline
\multirow{2}{10em}{EnFed} & 97\% (LSTM), & 98\% (LSTM),\\ 
& 96\% (MLP) & 98.05\% (MLP)\\
\hline
\multirow{2}{10em}{Cloud-only system without FL} & 97\% (LSTM),& 98\% (LSTM),\\ 
& 92\% (MLP)& 98\% (MLP)\\
\hline
    \end{tabular}
    \label{tab:base}
\end{table}
\subsection{Comparison with cloud-only framework}
\begin{figure}
\centering
    \includegraphics[width=0.9\linewidth, height=2.0in]{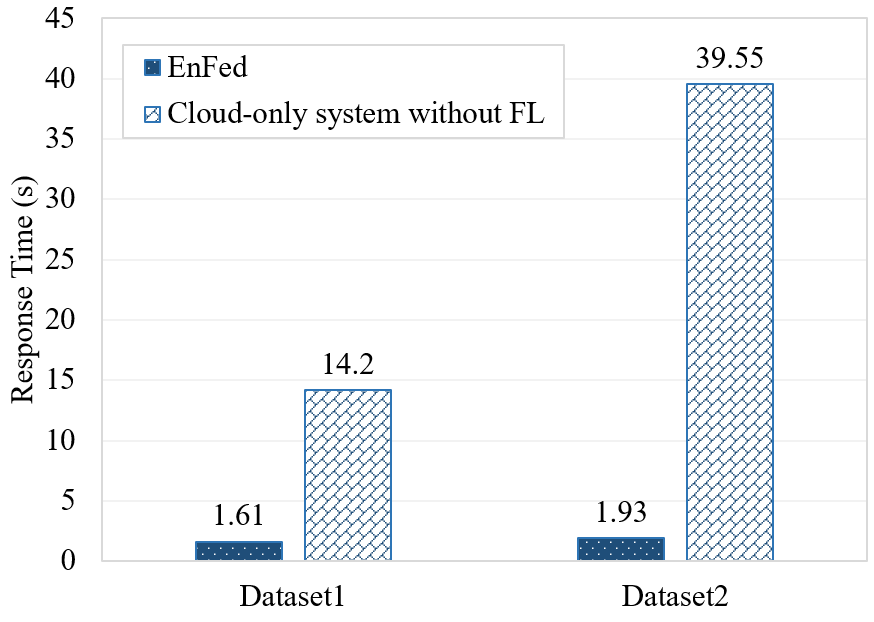}
    \caption{\centering{Response Time in EnFed and Cloud-only system using LSTM}}
    \label{c1}
\end{figure}

\begin{figure}
\centering
    \includegraphics[width=0.9\linewidth, height=2.0in]{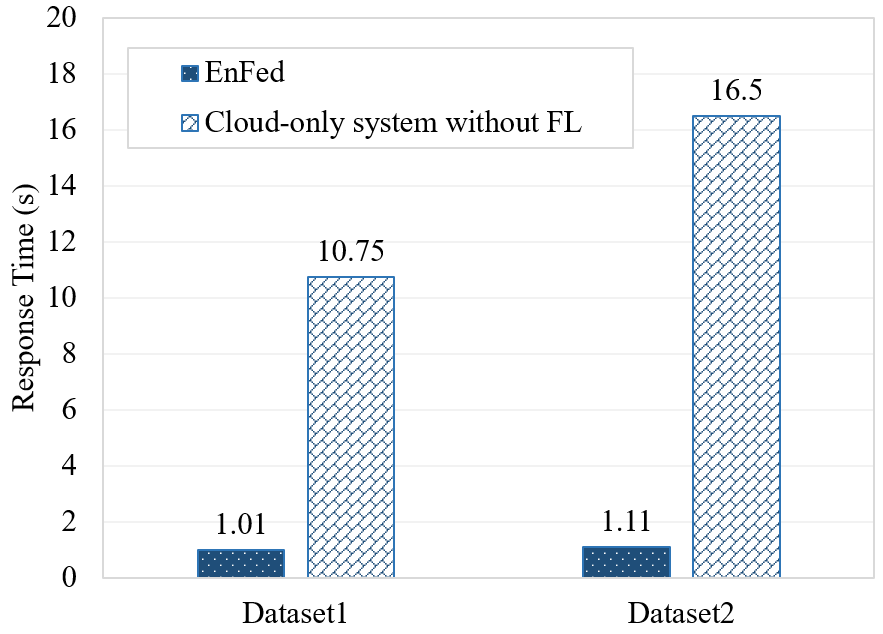}
    \caption{\centering{Response Time in EnFed and Cloud-only system using MLP}}
    \label{c2}
\end{figure}

Finally, we compare the response time and prediction accuracy of EnFed with the conventional cloud-only system without FL, where the user sends the data to the cloud for analysis and receives the result. We have conducted experiments using the first and second datasets and measured average response time and prediction accuracy. The average prediction accuracy for LSTM and MLP for the cloud-only system without FL are presented in Table \ref{tab:base} in comparison to EnFed. The average response time for EnFed and cloud-only system are presented in Figs. \ref{c1} and \ref{c2}. The average prediction accuracy for the first and second datasets using LSTM are 97\% and 98\%, respectively, for the cloud-only system without FL, and these values are the same as the prediction accuracy of EnFed using LSTM. The average prediction accuracy of the cloud-only system without FL for the first and second datasets are 92\% and 98\%, respectively, while using MLP. For the first dataset, EnFed has achieved 96\% prediction accuracy, and for the second dataset EnFed has a prediction accuracy of 98.05\%, using MLP, which is better than the cloud-only system. The response time for EnFed is approximately 89\% and 95\% less than the cloud-only system for the first and second datasets, respectively, while using LSTM. The response time for EnFed is approximately 91\% and 93\% less than the cloud-only system for the first and second datasets, respectively, while using MLP. Hence, we observe that EnFed outperforms the cloud-only system from the perspective of response time and overall prediction accuracy. 

\par
\textcolor{black}{We would like to mention that as the requesting device builds its model based on the model parameters received from the contributors, the trustworthiness and staleness of the models are important. In EnFed, the contributors only send their model updates with minimal loss. Hence, the model updates with minimal loss are aggregated to build the requesting device's model. However, there is a probability of model drift if the data distributions of the contributors' hugely differ from that of the requesting device. Further, the trustworthiness of the contributors' models is significant. Shannon's entropy \cite{lutz2025optimizing} can be used to select only trustworthy and reliable clients with low entropy and consistency. Furthermore, to  address the staleness issue, the requesting device can only select clients with the latest updated model parameters so that an accurate and stable model can be obtained.}  
\section{Conclusions and Future Work}
\label{con}
\textcolor{black}{Energy-efficient FL is a significant research area for sustainable mobile computing. The resource limitation and limited battery life of mobile devices are two important challenges for active participation in the CFL and DFL processes.} To address these issues, we have proposed an energy-efficient FL approach named \textit{EnFed} with a case study on the human activity recognition system. In the proposed approach, a device with poor network connectivity requests nearby devices for a model of an application with respect to an incentive. The nearby devices having updated models for that application accept the request if they agree to the offered incentive and send their model updates. The requesting device initializes and updates its model after receiving model updates from the contributors. \textcolor{black}{The proposed approach is applicable for a real-time environment where a mobile device with limited battery life, limited computational resources, and limited network connectivity requires an updated model for an application.} The performance of the proposed approach is evaluated with respect to the prediction accuracy, training time, energy consumption of the device, and response time. The experimental results show that EnFed has achieved above 95\% accuracy for the considered datasets while using LSTM and MLP. The experimental results also show that using EnFed the training time and energy consumption are reduced by approximately 59\% and 19\% than DFL for the first and second datasets, respectively, using LSTM. The results also show that EnFed has reduced the training time and energy consumption by approximately 85\% and 27\% compared to CFL for the first and second datasets, respectively, using LSTM. From the results, we also have observed that using EnFed with MLP, the training time and energy consumption are reduced by approximately 92\% for the first dataset and 96\% for the second dataset, compared to CFL using MLP. The results also present that EnFed with MLP has approximately 56\% and 73\% lower training time and energy consumption for the first and second datasets, respectively, than DFL using MLP. Finally, we conclude that EnFed maintains a trade-off between the prediction accuracy and the training time along with the energy consumption of the user device so that a model with good prediction accuracy is obtained at a lower time and lower energy consumption of the device. \par

\textcolor{black}{EnFed provides a personalized model for a device in a resource-constrained scenario by aggregating model updates from nearby devices and fitting it with its own dataset. However, the feature space of the nearby devices' datasets may differ from the feature space of the requesting device's dataset, or the feature space and the sample space both may differ for heterogeneous contributors. In the future, we would like to extend the scope of EnFed to address this challenge, along with the convergence analysis of the same. We would also like to explore the use of generative adversarial networks with FL to address the data scarcity and class imbalance issues. In the future, we would also like to use differential privacy mechanisms in EnFed for lightweight privacy management. The use of Bayesian LSTM in FL-based activity recognition is another future research direction of EnFed.}\\

\section*{Acknowledgements}
This work is partially supported by ARC Discovery Project (DP240102088) and ARC LIEF grants.

\bibliography{ref}
\bibliographystyle{IEEEtran}

\section*{Biography}
\begin{IEEEbiographynophoto}
{\textbf{Anwesha Mukherjee} is an Assistant Professor of the Department of Computer Science, Mahishadal Raj College, West Bengal, India. She is a research visitor in the qCLOUDS Lab, The University of Melbourne.} 
\end{IEEEbiographynophoto}
\vspace{-250pt}
\begin{IEEEbiographynophoto}
{\textbf{Rajkumar Buyya} (Fellow, ACM; Fellow, IEEE) is Director of the Quantum Cloud Computing and Distributed Systems (qCLOUDS) Laboratory, The University of Melbourne, Australia. He has authored more than 850 publications and seven text books including “Mastering Cloud Computing” published by McGraw Hill, China Machine Press, and Morgan Kaufmann for Indian, Chinese and international markets respectively. He is one of the highly cited authors in computer science and software engineering worldwide (h-index=180, g-index=384, 169,500+ citations). }
\end{IEEEbiographynophoto}

\end{document}